\documentclass[aps,prb,reprint,unsortedaddress, showkeys]{revtex4-2}

\usepackage{graphicx}
\usepackage{bm}
\usepackage{amsmath}
\usepackage{xspace}
\usepackage{overpic}

\newcommand{\p}{\mathcal{P}}
\newcommand{\data}{\mathcal{D}}
\newcommand{\model}{\mathcal{M}}
\newcommand{\like}{\mathcal{L}}
\newcommand{\basis}{\mathcal{B}}

\newcommand{\bmx}{\bm{x}}

\newcommand{\bmw}{\bm{w}}

\newcommand{\bmI}{\bm{I}}

\newcommand{\bmK}{\bm{K}}
\newcommand{\bmC}{\bm{C}}
\newcommand{\bmS}{\bm{\Sigma}}
\newcommand{\bmY}{\bm{Y}}
\newcommand{\dm}{\bm{\Phi}}

\newcommand{\bmA}{\bm{A}}
\newcommand{\gdist}{\mathcal{N}}

\newcommand{\cost}{\mathcal{C}}
\newcommand{\costb}{\cost^\text{MSE}}
\let\oldput\put
\def\put(#1,#2)#3{%
  \oldput(#1,#2){\sffamily #3}%
}

\begin{document}

\title{Uncertainty-aware electronic density-functional distributions.}

\author{Teitur Hansen}
\email{teih@dtu.dk}
\affiliation{Department of Physics, Technical University of Denmark}

\author{Jens Jørgen Mortensen}
\email{jjmo@dtu.dk}
\affiliation{Department of Physics, Technical University of Denmark}

\author{Thomas Bligaard}
\email{tbli@dtu.dk}
\affiliation{Department of Energy Conversion and Storage, Technical University of Denmark}

\author{Karsten Wedel Jacobsen}
\email{kwja@dtu.dk}
\affiliation{Department of Physics, Technical University of Denmark}

\date{\today}

\begin{abstract}
We introduce a method for the estimation of uncertainties in density-functional-theory (DFT) calculations for atomistic systems. The method is based on the construction of an uncertainty-aware functional distribution (UAFD) in a space spanned by a few different exchange-correlation functionals and is illustrated at the level of generalized-gradient-approximation functionals. The UAFD provides reliable estimates of errors -- compared to experiments or higher-quality calculations -- in calculations performed self-consistently with the Perdew-Burke-Ernzerhof functional. The scheme furthermore allows for a decomposition of the error into a systematic bias and a reduced error. The approach is applied to four different properties: molecular atomization energies, cohesive energies, lattice constants, and bulk moduli of solids. The probability distribution can be tailored to optimize the prediction of a single property or for several properties simultaneously.
\end{abstract}

\maketitle

\section{Introduction}
Density functional theory (DFT) is one of the most widely used computational techniques to describe materials and/or molecules at the electronic scale \cite{hohenberg_inhomogeneous_1964, kohn_self-consistent_1965}. With currently more than 90 scientific publications per day using the approach \cite{clarivate2025wos}, the impact of the theory in the fields of chemistry and materials science is considerable. Although DFT is formally exact, various aspects contribute inaccuracies to DFT simulations. Some of these error contributions, such as those that stem from the numerical representation of electron orbitals, densities, potentials, and sampling of k-points, can be systematically converged \cite{gabrielUncertaintyQuantificationMaterials2020}. Other error contributions, such as those originating from an approximate treatment of core electrons and relativistic effects, can be limited by careful benchmarking and comparisons between different implementations \cite{PizziNRP2024}. A challenging remaining error contribution in DFT simulations is the exchange-correlation functional, which, although in principle well defined, needs approximations, which have been classified into a number of levels \cite{perdew_jacobs_2001} according to accuracy and complexity. The development of large simulated material property databases in the past decade \cite{materialsproject,oqmd} has led to a renewed focus on reducing and estimating errors in DFT simulations. Databases combining simulated and experimental data allow for regressing physically informed statistical models \cite{amandawang2021,pernotPredictionUncertaintyDensity2015}, which typically perform well but must be trained for each individual material property. Atomistic machine learning models trained on simulations have led to a wealth of error estimation methods \cite{errorReview2024} through the use of e.g. bootstrapping \cite{rune2017, proppeReliableEstimationPrediction2017}, Gaussian processes \cite{tranMethodsComparingUncertainty2020,liuHighthroughputHybridfunctionalDFT2024}, Monte Carlo dropout \cite{MCdropout}, conformal prediction \cite{conformalPrediction}, Bayesian neural networks \cite{pathrudkarElectronicStructurePrediction2024}, and neural network ensembles \cite{ceriottiNNensemble}.

\section{Definition of functional distribution}
Here, we develop an approach to estimate the accuracy of DFT calculations based on probability distributions of exchange-correlation functionals (xc-functionals). We consider a space of functionals, $\model$, which is described by a set of parameters $\bmw$ so that a given value of $\bmw$ corresponds to a choice of xc-functional. In this space, we consider a probability distribution $\p_\model(\bmw)$ to be determined in the following.

For a particular atomic system, $\bmx$, defined by the chemical elements of the atoms and their positions, and for a particular property, $y$, the functional corresponding to $\bmw$ provides a prediction, which we denote by $y(\bmx,\bmw)$. The probability distribution in model space, thus leads to a distribution of predictions of $y$ through
\begin{equation}\label{eq:predict}
    \p_p(z|\bmx) = \int \delta(z-y(\bmx,\bmw))\p_\model(\bmw)\, d\bmw.
\end{equation}

To determine the probability distribution, we introduce a set of accurate reference data (from experiments or results from converged quantum chemical calculations) consisting of atomic systems, $\bmx_n$, with given reference (``target'') values of properties, $t_n$, for a collection of systems ($n=1,2,\ldots,N$).
What we then propose is, to determine the probability distribution, $\p_\model$, by a direct optimization of the likelihood
\begin{equation}\label{eq:L}
    \like[\p_\model] = \prod_n^N \p_p(t_n|\bmx_n),
\end{equation}
which involves the distribution in model space through Eq.~(\ref{eq:predict}). A similar likelihood has been used for uncertainty estimation using deep ensembles \cite{lakshminarayanan_simple_2017, busk_calibrated_2022}. The optimization of Eq.~(\ref{eq:L}) leads to an ``uncertainty-aware'' functional distribution (UAFD), which favors functionals with predictions close to the experimental data, but with a sufficient width to provide realistic uncertainty estimates. 

\begin{figure}
\includegraphics[width=0.8\linewidth]{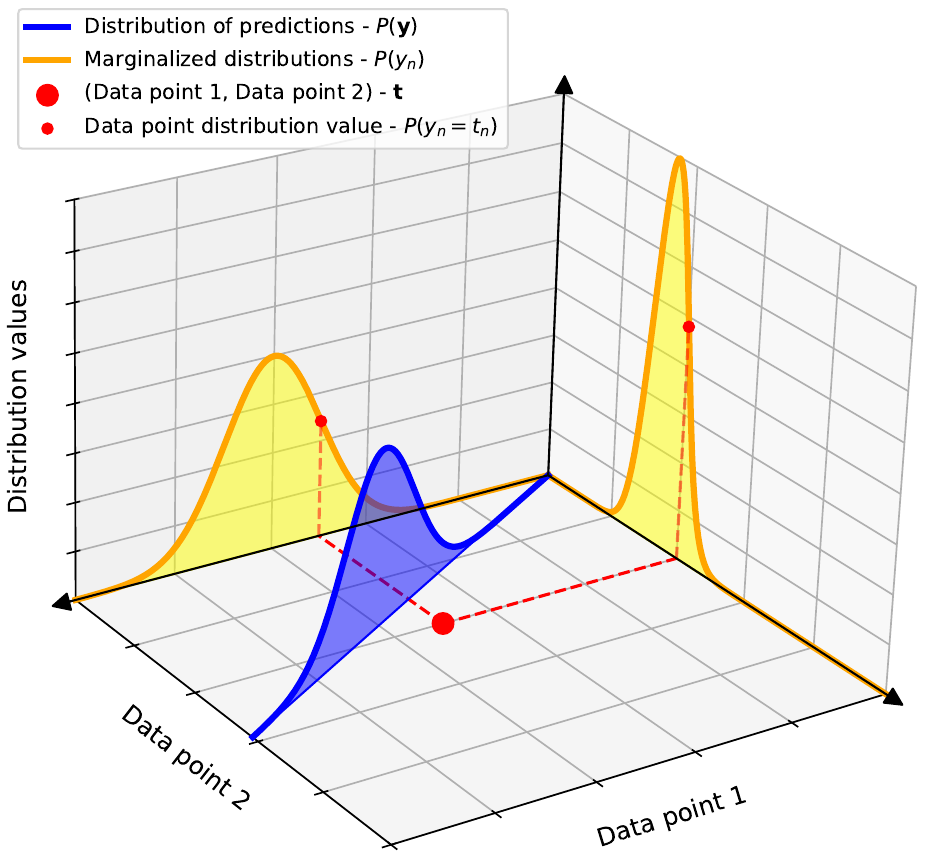}
\caption{An example of a one-dimensional model space and two data points with Gaussian distributions. The blue curve represents the probability distribution of the model predictions, while the yellow curves are the marginal distributions. Optimization of $C(\bmw_0,\bmK)$ corresponds to modifying the blue distribution to maximize the product of the probabilities at the two red points on the yellow distributions.}
\label{fig:data_marginalization}
\end{figure}

To illustrate the approach, we show in Fig.~\ref{fig:data_marginalization} an example with a one-dimensional model space with a Gaussian distribution and two data points. If we assume a linear relation between the model parameter and the predicted data values, we obtain a Gaussian prediction in the data space (blue curve), which results in two Gaussian marginal distributions for the data (yellow curves). No model can reproduce both data points (the red point in the horizontal plane) because the blue line is fixed by the constrained model space. The optimization in Eq.~(\ref{eq:L}) corresponds to maximizing the product of the prediction probabilities of the two data points (the red points on the yellow distributions). A high value is obtained if the two red points are close to the top of narrow distributions, but if this is not possible, the distribution in model space broadens and constitutes a compromise between prediction values and uncertainties.

We now proceed to show how this is implemented for a model space, $\model$, consisting of linear combinations of a set of functionals, $\basis$. We choose a linear model for the energy, $E(\bmx) = \sum_i w_i E_i(\bmx),\,i\in\basis$, in order to obey scaling with the system size, and, more generally, we shall assume that the considered properties, $y$, can be approximately obtained by linear interpolation $y(\bmx,\bmw) = \sum_i \phi_i(\bmx) w_i$, where $\phi_i(\bmx)$ is the value obtained with functional $i$. If we assume a Gaussian distribution, $\p_\model(\bmw) = \gdist(\bmw|\bmw_0,\bmK)$, with mean $\bmw_0$ and covariance $\bmK$, the predictive distribution Eq.~(\ref{eq:predict}) for a data point $(\bmx_n, t_n)$ also becomes Gaussian with mean $\bar{y}_n = (\dm\bmw_0)_n$ and variance $\sigma_n^2 = (\dm \bmK \dm^T)_{nn}$, where we have introduced the so-called design matrix $\dm_{ni} = \phi_i(\bmx_n)$ \cite{bishop_pattern_2006}. The negative log of the likelihood in Eq.~(\ref{eq:L}) can then be written
\begin{align}\label{eq:loglike}
\begin{split}
    \cost(\bmw_0,\bmK) := &-\log(\like)\\
    = &\frac{1}{2}\sum_n (t_n-\bar{y}_n)^2/\sigma_n^2+ \frac{1}{2}\sum_n \log(\sigma_n^2)\\ &\hspace{3cm} + \frac{N}{2}\log(2\pi),
\end{split}
\end{align}
which is an effective cost function that should be minimized to obtain $\bmw_0$ and $\bmK$.

We see that in the cost function each data point has a natural weight given by the uncertainty parameter $\sigma_n$. This leads to some very favorable features: 1) The cost function is independent of scaling, \emph{i.e.} if the size of a unit cell in a periodic system is doubled and the energies also increase by a factor of two, the corresponding term in the cost function is unchanged. It is for example also independent of whether an atomization energy is given per atom or per molecule. 2) The cost function is uniquely defined also for inhomogeneous data with, for example, different units (such as cohesive energies and lattice constants). The noise parameters make the terms in the cost function dimensionless. 3) The variances provide a natural weighting of individual data points within a dataset. For a given space of functionals, the predictions of a given property might be consistently better for one class of systems than for another, leading to a natural different weighting of data points in the two classes. We shall see an example of this for atomization energies, where it turns out that predictions become more accurate for hydrocarbons than for other molecules.

\section{Comparison with traditional Bayesian approach}
Before we proceed, we would like to compare our method with a traditional Bayesian analysis and discuss why this is not well-suited for our context. In the traditional Bayesian approach, the posterior distribution, $\p_{\text{Bayes}}$ is given by the likelihood and the prior distribution as
\begin{equation}\label{eq:bayes}
    \p_{\text{Bayes}}(\bmw) \propto \prod_n \gdist(t_n|y(x_n,\bmw),\sigma^2) \p_0(\bmw),
\end{equation}
where the likelihood is taken as a Gaussian distribution of the data around the model prediction with noise $\sigma$. Setting the prior to one, the optimization of the posterior distribution corresponds to the minimization of the mean-squared-error (MSE) cost function
\begin{equation}\label{eq:bayes_cost}
    \costb(\bmw) = \sum_n (t_n - y(x_n,\bmw))^2.
\end{equation}

There are several issues with this approach in our context. The main reason why the predictions do not reproduce the data is not because of noise in the data or lack of precision in the calculations, but because of the incompleteness of the model space. For example, no GGA can predict molecular atomization energies with an accuracy less than the errors in high-level quantum chemistry calculations \cite{curtiss_assessment_2000}. The Bayesian approach implicitly assumes that the correct model is included in the model space (within the noise level), which is not our situation. This point is emphasized by the fact that as more data points are added, the distribution in model space as given by Eq.~(\ref{eq:bayes}) becomes more narrow, leading to smaller and smaller uncertainties in the predictions based on Eq.~(\ref{eq:predict}). This is not the correct behavior when the errors are due to a basic incompleteness of the model space. 

Despite these issues, Eq.~(\ref{eq:bayes}) has been used with some success to generate ensembles of interatomic potentials \cite{brown_statistical_2003,frederiksen_bayesian_2004,kurniawan_bayesian_2022} and also the so-called BEEF electronic exchange-correlation functional ensembles with error estimation \cite{mortensen_bayesian_2005, wellendorff_density_2012, wellendorff_mbeef_2014, lundgaard_mbeef-vdw_2016, medford_assessing_2014}. These applications involve a pragmatic rescaling of the noise parameter to counteract the collapse of the uncertainties as more data points are added. We also note that the three advantageous features listed above for the cost function $\cost$ do not hold for the MSE cost function.

The traditional Bayesian approach allows for a broader interpretation of the noise parameter appearing in the likelihood in Eq.~(\ref{eq:bayes}). If this parameter is optimized based on for example the marginalized likelihood or marginalized posterior, it will not only represent the actual noise on the data, but also the deviation between the mean model and the data even if this deviation is due to incompleteness of the functional space. In the limit of large amounts of data, the distribution in model space still becomes narrow leading to small uncertainties from the model distribution. However, if the predictive distribution includes both the model uncertainty and the likelihood including the (now re-interpreted) noise, a final prediction uncertainty is obtained also in the large-data limit. This approach has been used by Aldegunde et al \cite{aldegunde_development_2016} (together with some further refinements in form of prior distributions) to develop a functional with uncertainty quantification. Our approach is fundamentally different by the use of the likelihood Eq.~(\ref{eq:L}). This likelihood involves the full parameter distribution $\gdist(\bmw|\bmw_0,\bmK)$ simultaneously and cannot be reduced to a likelihood for each value of the parameters $\bmw$ as the likelihood $\prod_n \gdist(t_n|y(x_n,\bmw),\sigma^2)$ in Eq.~(\ref{eq:bayes}). Therefore, the errors on predictions become described by fluctuations in model space, which prevail also in the limit of large amounts of data and no noise.

\section{Regularization}
The cost function, $\cost$, has a divergence issue similar to Gaussian mixture models \cite{bishop_pattern_2006}. If the probability distribution concentrates around a particular data point with $\bar{y}_n = t_n$, the variance $\sigma_n^2$ can vanish, leading to a (negative) divergence of the term $\log(\sigma_n^2)$. We address this issue, as well as potential overfitting, by two types of regularization. The first is to associate a width to the value of $\bmw_0$ of the form $\gdist(\bmw_0|\bar{\bmw}_0,\lambda_K \bmI)$ with a new parameter $\lambda_K$. This leads to a new distribution $\p_\model(\bmw) = \int \gdist(\bmw|\bmw_0,\bmK) \gdist(\bmw_0|\bar{\bmw}_0,\lambda_K) d\bmw_0 = \gdist(\bmw|\bar{\bmw}_0,\bmK+\lambda_K)$, where the effect is to add $\lambda_K$ to the diagonal of $\bmK$. (In the following, we denote the new mean, $\bar{\bmw}_0$, by just $\bmw_0$.)

The second regularization, which counteracts overfitting, consists in adding a term $-\lambda_S S$ to the cost function, where $\lambda_S$ is a constant, and $S$ is the entropy
\begin{align}
S &= - \int \p_\model(\bmw)\log(\p_\model(\bmw))\, d\bmw\ \nonumber \\
  &= \frac{M}{2}\log(2\pi e) + \frac{1}{2}\log(\det(\bmK+\lambda_K)).\label{eq:entropy}
\end{align}
The values of the regularization parameters $\lambda_K$ and $\lambda_S$ are determined by cross-validation. The data is split into 80\% training and 20\% validation in five different cases such that 100\% of the data has been validated on. We determine the lowest cost function value by performing a grid search in the $(\lambda_S,\lambda_K)$-space. We average over 10 random orderings of the data to avoid dependence on splitting of the training and validation sets. 

The resulting cost function, $\cost(\bmw_0,\bmK+\lambda_K)-\lambda_S S$, can now be minimized. It is quadratic in $\bmw_0$, which can therefore be determined analytically. The derivative of the cost with respect to $\bmK$ can also be obtained analytically as shown in the Appendix.

\section{Model space}
We consider a model space spanned by four GGAs and LDA, $\basis = \{\text{PBE, RPBE, BLYP, PBEsol, LDA}\}$ \cite{LDA,PBE,PBEsol,RPBE,BLYP1,BLYP2}, where the calculations are performed self-consistently with PBE, and the calculations with the other functionals are performed non-self-consistently based on the PBE density.
We label the different functionals in the mentioned order with an index $i = 0,1,2,3,4$.

We use non-self-consistent evaluations of the functionals because of the computational advantage. Once the PBE electronic density is known for a particular system, the evaluation of the other GGA functionals require only simple integrals over space. The use of non-self-consistent functionals for the functional distribution is in principle correct in so far as the same approach is used for learning the functional distribution and for its application. However, the non-self-consistent functional distribution is in fact also very close to the self-consistent one as discussed further below when considering the datasets. It is also possible to work directly with a distribution based on self-consistent functionals and that could be required when considering more complicated properties involving the self-consistent calculation of atomic forces.

We use a sum rule for the linear combination of xc-functionals that the coefficients should add up to one. If we write the energy, $E(\bmx)$, for a given system, $\bmx$, as
\begin{equation}
    E(\bmx)=\sum_{i=0}^4 w_i E_i(\bmx),
\end{equation}
where $i$ runs over the different functionals, we require $\sum_i w_i = 1$. In practice this is achieved by subtracting the PBE result (denoted by $E_0$) from all energies, and setting $w_0 = 1-\sum_{i=1}^4 w_i$. We then work with one parameter less in the calculations with the energy expression
\begin{align}
    E(\bmx)-E_0(\bmx) & = \sum_{i=0}^4 w_i (E_i(\bmx)-E_0(\bmx))\\
    &=\sum_{i=1}^4 w_i (E_i(\bmx)-E_0(\bmx)).
\end{align}

\section{Prediction}

The minimization of the cost function determines the values of $\bmw_0$ and $\tilde{\bmK} = \bmK + \lambda_K$. The UAFD is then given by $\p_\model(\bmw) = \gdist(\bmw|\bmw_0,\tilde{\bmK})$, where $\tilde{\bmK} = \bmK + \lambda_K$.

For a given property, $y$, the calculation by the five functionals for a system $\bmx$ is denoted $\phi_i(\bmx)$. The average prediction, $\bar{y}(\bmx)$, is then given by 
\begin{equation}
    \bar{y}(\bmx) = \phi_\text{PBE}(\bmx) + \sum_{i=1}^4 w_{0,i} (\phi_i(\bmx) - \phi_\text{PBE}(\bmx)).
\end{equation} 
The variance around the average prediction is determined by
\begin{equation}
    \sigma^2(\bmx) = \sum_{i=1}^4\sum_{j=1}^4 
    (\phi_i(\bmx)-\phi_\text{PBE}(\bmx))
    \tilde{K}_{ij}
    (\phi_j(\bmx)-\phi_\text{PBE}(\bmx))
\end{equation}

\section{Datasets}
There are four datasets used for training and validation. The first is a dataset taken from the BEEF-vdW project \cite{wellendorff_density_2012} and contains 222 atomization energies. The calculated energies are compared with the reference atomization energies from G3/99 \cite{curtiss_assessment_2000}. The calculations are done using plane-waves and the projector-augmented wave method as implemented in the GPAW electronic structure code\cite{mortensen_real-space_2005, mortensen_gpaw_2024, hjorth_larsen_atomic_2017}.

The other three datasets are cohesive energies, equilibrium lattice constants, and bulk moduli of 44 solids as calculated for different exchange-correlation functionals by Tran et al. \cite{tran_rungs_2016}. The equilibrium lattice constants and bulk moduli are obtained by fitting to calculations with varying lattice constants \cite{tran_rungs_2016}.

All four datasets are based on atomic structures optimized with PBE, and the results with the other functionals are calculated non-self-consistently based on the PBE electronic density. The non-self-consistent energies are in fact quite close to the self-consistent ones as validated by performing the self-consistent calculations. For example, the mean absolute error on the LDA atomization energies due to non-self-consistency is only 0.005 eV/atom, while, for comparison, the mean absolute difference between the LDA and the PBE atomization energies is 0.52 eV/atom.

\begin{figure}[t!]
\begin{overpic}[width=0.8\linewidth]{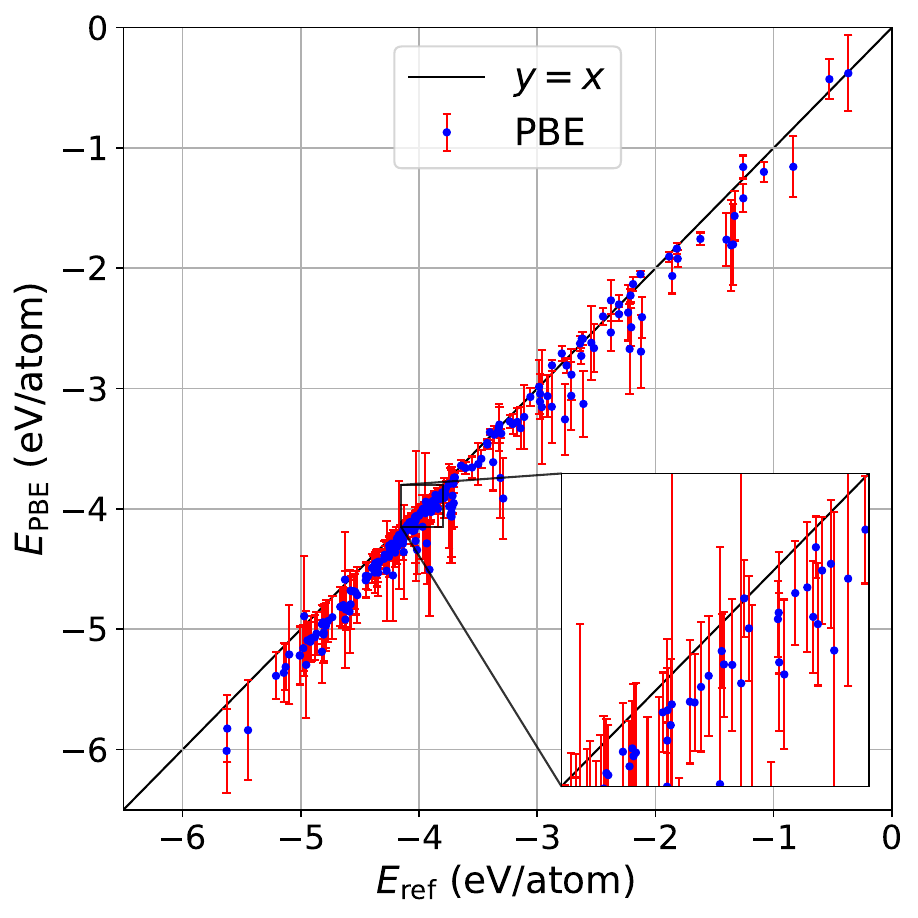}
\put(15,90){\large a)}
\end{overpic}
\begin{overpic}[width=0.8\linewidth]{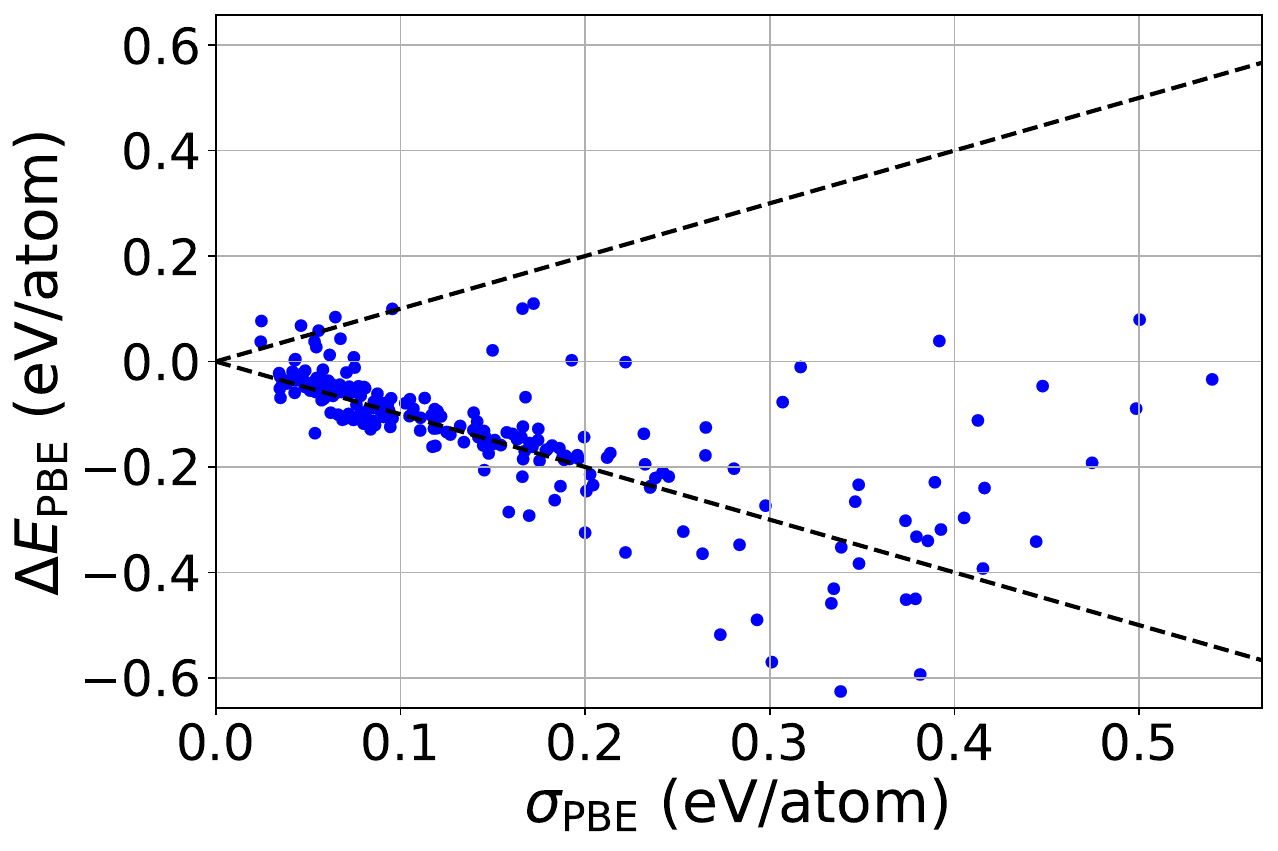}
\put(21,58){\large b)}
\end{overpic}
\begin{overpic}[width=0.8\linewidth]{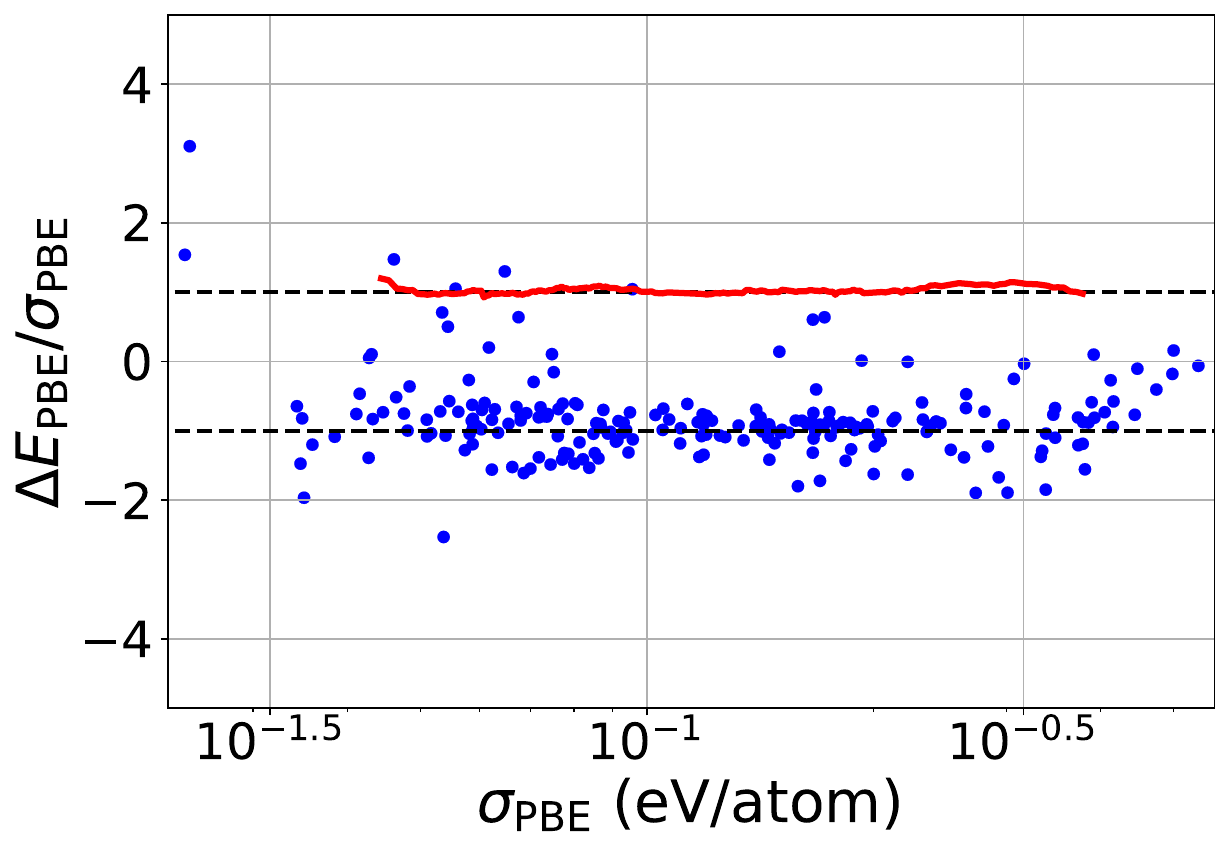}
\put(15,60){\large c)}
\end{overpic}
\caption{a) Comparison between G3/99 and PBE-calculated atomization energies per atom including UAFD-uncertainties. b) Comparison between the estimated uncertainties of the PBE-calculated atomization energies and the actual errors. It is seen that the main reason for the error is the systematic overbinding of PBE.
c) The ratio of the error relative to the predicted uncertainty (the normalized error) as a function of the predicted uncertainty. The red curve shows a moving root-mean-square value of the normalized error (RMSNE) over $N_a=30$ data points as explained in detail in the text. The fact that the RMSNE is close to one is an indication that the uncertainty estimates are appropriate.}
\label{fig:atomization_PBE}
\end{figure}

\section{Results}

We first consider a single dataset, $\data_\text{atom}$, consisting of $N_\text{atom} = 222$ molecular atomization energies. The hyperparameters are determined as described above to be $(\lambda_S,\lambda_K)=(2\cdot10^{-2},10^{-6})$.

The most basic question to ask is whether the results of the five functionals do contain sufficient information to estimate uncertainties. This is addressed in Fig.~\ref{fig:atomization_PBE}a), where we show a comparison between the atomization energies calculated with PBE and the experimental values together with the uncertainties $\sigma_\text{PBE}$ determined by the UAFD. The uncertainties are given by the probability distribution Eq.~(\ref{eq:predict}) as the variance around the PBE value $\sigma_\text{PBE}^2 = \int (z-y_\text{PBE}(\bmx))^2 \p_p(z|\bmx)\,dz$, where $y_\text{PBE}(\bmx)$ is the prediction by the PBE functional for the system given by $\bmx$. We use five-fold cross-validation, so that the dataset is split in five, where 4/5 of the dataset is used for training, and the remaining 1/5 is used for testing. It is the test results, which are shown in Fig.~\ref{fig:atomization_PBE}. The error bars are seen to generally reach from the PBE values to the experimental values.

We also show a correlation plot between the predicted uncertainties and the actual errors in Fig.~\ref{fig:atomization_PBE}b). The errors, which are mostly due to the overbinding by PBE, are clearly well estimated by the calculated uncertainties.

To quantitatively assess the uncertainties, we introduce a normalized error as the difference between PBE and the reference values (\emph{i.e.} the errors) divided by the predicted uncertainties.
Fig~\ref{fig:atomization_PBE}c) shows the normalized errors as a function of the uncertainties now on a logarithmic scale.

We also show a moving root-mean-square value of the normalized errors (RMSNE), which is defined in the following way: The data points are sorted by increasing predicted uncertainties so that $\sigma_n <= \sigma_{n+1}, n = 1,2, \ldots, N$. The actual error for a data point is denoted $\Delta E_n$. For each $j = 1,2,\ldots, N-N_a$, where $N_a$ is an integer indicating the number of points in the averaging, we now calculate the RMSNE:
    \begin{equation}
        \text{RMSNE}_j = \sqrt{\frac{1}{N_a}\sum_{i=j}^{j+N_a-1} \left(\frac{\Delta E_i}{\sigma_i}\right)^2}.
    \end{equation}
This value is then shown in the plot at the coordinates $(\bar{\sigma}_j, \text{RMSNE}_j)$, where $\bar{\sigma_j}$ indicates the mean value of the predicted error in the averaging window $\bar{\sigma}_j = \sum_{i = j}^{j+N_a-1} \sigma_i/N_a$. We use an averaging window of $N_a = 30$, and the points in the plot are connected to a red curve.

The RMSNE value should be close to one for a successful error prediction and this is clearly the case as seen from the red curve in Fig.~\ref{fig:atomization_PBE}c). The plot also confirms that most of the error is due to a systematic overbinding by PBE, as can be seen from the points in Fig.~\ref{fig:atomization_PBE}b) being scattered around minus one.

\begin{figure}[t!]
\begin{overpic}[width=0.8\linewidth]{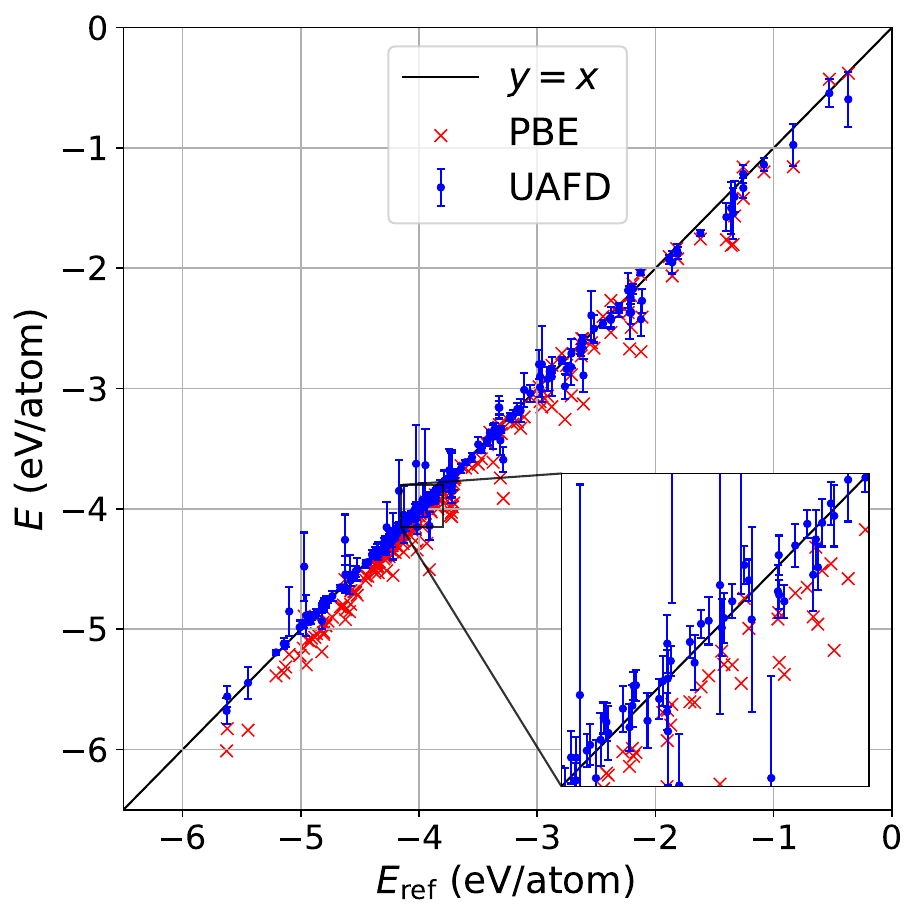}
\put(15,88.5){\large a)}
\end{overpic}
\begin{overpic}[width=0.8\linewidth]{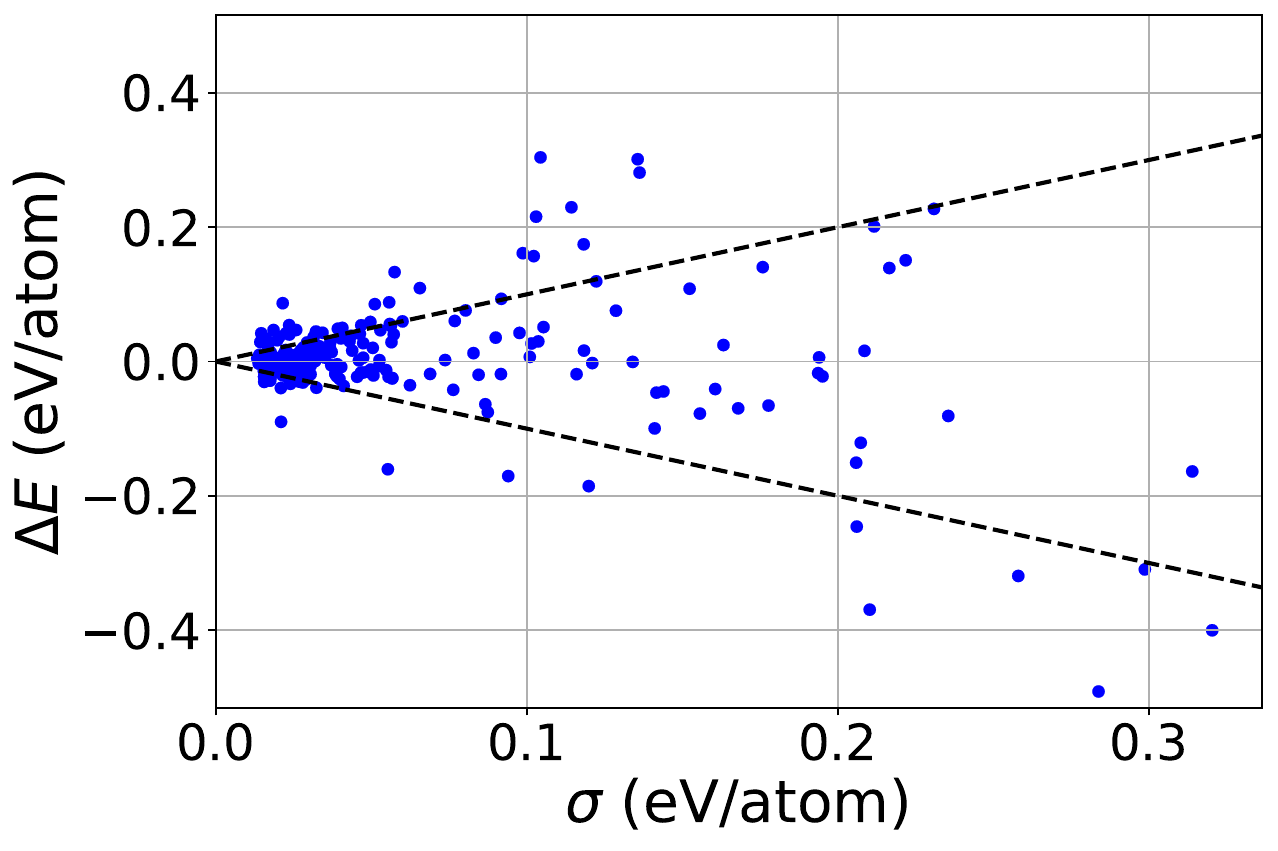}
\put(19,58.5){\large b)}
\end{overpic}
\begin{overpic}[width=0.8\linewidth]{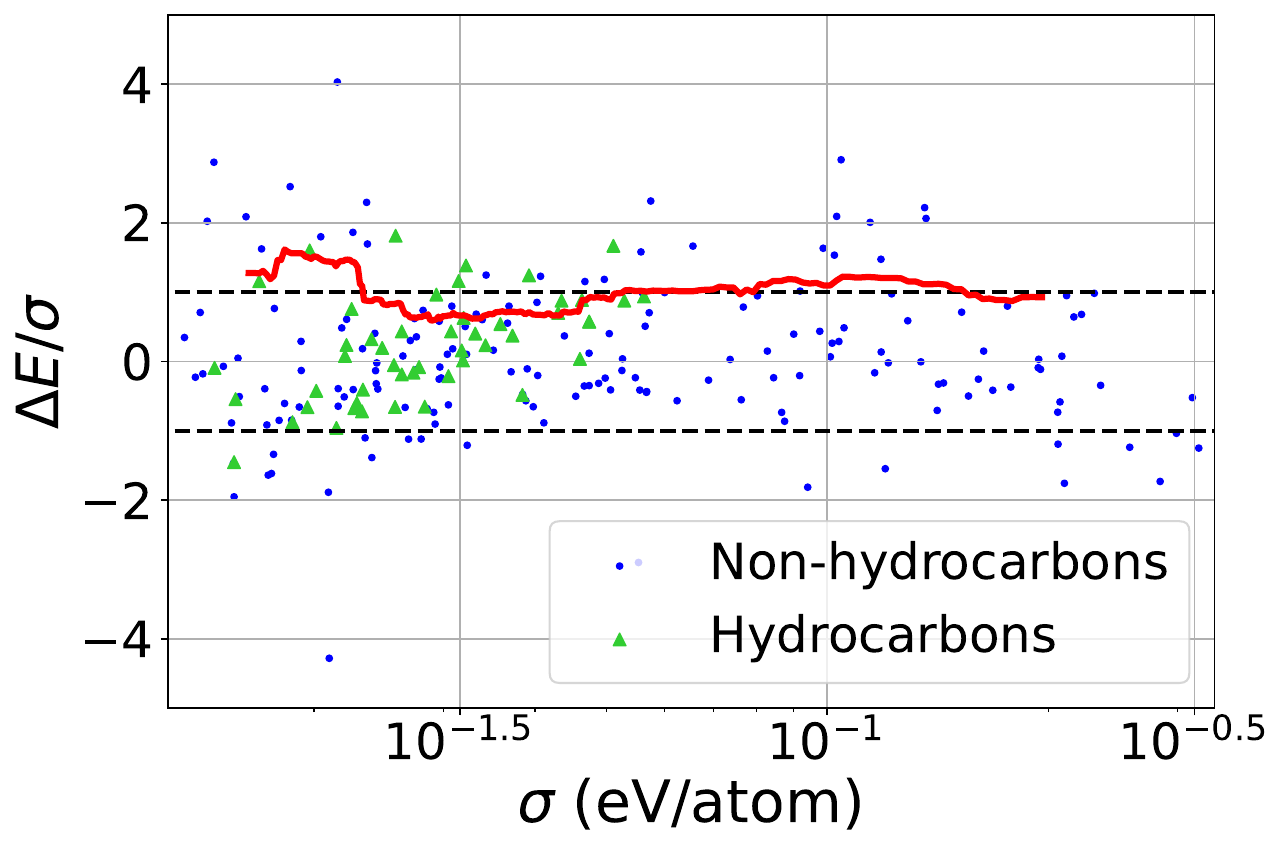}
\put(15,58.5){\large c)}
\end{overpic}
\caption{a) Comparison between experimental and calculated atomization energies per atom including uncertainties from UAFD. The average model is seen to correct for the systematic over-binding of PBE. b) A correlation plot of the estimated uncertainty and the actual error for the average model.
c) The ratio of the error relative to the predicted uncertainty as a function of the predicted uncertainty. The points with green triangles are hydrocarbons, while the blue circles are the rest of the molecules. The red curve shows a moving RMSNE value over $N_a=30$ data points of the normalized error. The uncertainty estimates are seen to vary by more than one order of magnitude with the hydrocarbons exhibiting relatively small uncertainties.}
\label{fig:atomization}
\end{figure}

The approach allows for a decomposition of the PBE uncertainty, so that we can in fact remove the systematic overbinding and make improved predictions with smaller errors. The probability distribution defines an average model (\emph{i.e.} an average xc-functional), $\bar{y}(\bmx) = \sum_i w_{0,i} \phi_i(\bmx)$, and the PBE uncertainty estimate is given by $\sigma_\text{PBE}(\bmx) = \sqrt{(y_\text{PBE}(\bmx)-\bar{y}(\bmx))^2+\sigma(\bmx)^2}$, where $y_\text{PBE}(\bmx)$ denotes the prediction by PBE, and the uncertainty for the average model is given by $\sigma(\bmx)^2 = \sum_{ij} \phi_i(\bmx)K_{ij}\phi_j(\bmx)$. (Practical details of the calculations are shown in the Appendix.)

Fig.~\ref{fig:atomization} shows results similar to Fig.~\ref{fig:atomization_PBE}, but where the average model $\bar{y}$ is used for prediction instead of PBE. As can be seen by the more symmetric distribution of points around the $x$-axis in Figs.~\ref{fig:atomization}b) and \ref{fig:atomization}c), the systematic overbinding of PBE has been removed. The RMSE of the predictions are therefore decreased from 0.179 eV/atom by PBE to 0.090 eV/atom by the average xc-functional. The average functional is seen to predict hydrocarbons with particularly low errors, an example of how training with the cost function Eq.~(\ref{eq:loglike}) can lead to physically informed functionals. The natural weighting of the data points in the cost by the uncertainty makes it possible for the average model to distinguish between different types of data points. The error estimation is quite reasonable as shown by the (red) RMSNE curve in Fig.~\ref{fig:atomization}c) being close to one.

\begin{figure}
\begin{overpic}[width=\linewidth]{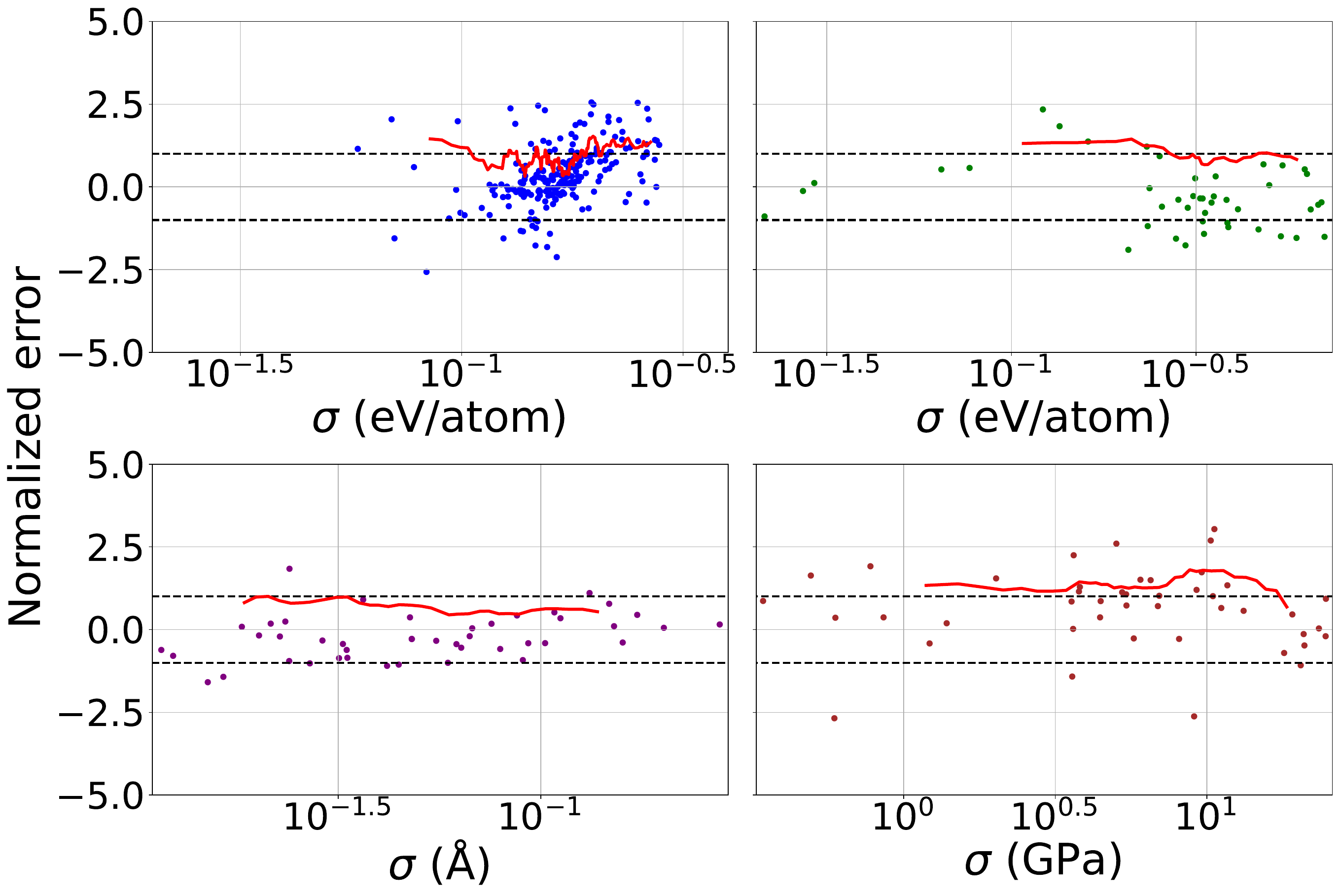}
\put(12.5,62){\footnotesize Atomization energies}
\put(57.5,62){\footnotesize Cohesive energies}
\put(12.5,29){\footnotesize Lattice constants}
\put(57.5,29){\footnotesize Bulk moduli}
\end{overpic}
\caption{Uncertainty estimates for a model trained on four different properties simultaneously. A moving RMSNE value over $N_a=10$ points has been used to evaluate the uncertainty prediction. It can be seen that the moving RMSNE value hovers around 1 of the normalized error and therefore is a decent error estimate. All the data have been 5-fold cross-validated.}
\label{fig:Uncertainty_estimate_4db}
\end{figure}

We now apply the approach to several different datasets simultaneously. We consider the dataset of atomization energies used above together with three properties of 44 bulk materials: cohesive energy, lattice constant, and bulk modulus \cite{tran_rungs_2016}. As discussed above, the cost function (Eq.~\ref{eq:loglike}) has a natural weighting of each data point given by the uncertainty. However, in the present case the datasets are of rather different size, and to obtain a well-balanced model, we introduce an additional weight factor, $W_\alpha$ for the points in dataset $\alpha$. It is given by $W_\alpha =1/\sum_\beta (N_\alpha/N_\beta)$, where $N_\alpha$ denotes the number of data points in set $\alpha$, so that the sets appear with the same weight in the cost. The hyperparameters are determined as described above to be $(\lambda_S,\lambda_K)=(10^{-2},10^{-6})$.

Optimizing the functional ensemble (including the five-fold cross-validation) leads to the results shown in Fig.~\ref{fig:Uncertainty_estimate_4db}. For all four properties, the ensemble provides reasonable uncertainties, as indicated by the red curves being close to one. We note that the error estimates for the atomization energies are spread over a smaller range than in the case of the atomization dataset alone (Fig.~\ref{fig:atomization}). This is due to the necessary compromise in the GGA xc-functional space between functionals that work well for molecules and those that work well for solids \cite{mortensen_bayesian_2005}. If we do not introduce the weighting factors $W_\alpha$ so that all data points have the same weight, the range of uncertainties for the atomization energies is broad like in Fig.~\ref{fig:atomization}, while the errors on, for example, the cohesive energies are somewhat larger because of the lower weight on this dataset.

In conclusion, we have established a method to construct xc-functional probability distributions, where the fluctuations provide realistic uncertainty estimates. 
The distribution can be tailored for a single property, as exemplified by the atomization energies, leading to a wide distribution of uncertainties. It is also possible to generate more widely applicable ensembles based on several different properties. The examples shown here are for a simple five-dimensional LDA/GGA space, but the approach should also be possible at higher levels of xc-approximations. The method as presented here is appropriate in the limit where calculations are precise and noise on the data can be neglected. Further investigations will show to what extent noise can be incorporated in the approach.

\begin{acknowledgments}
We acknowledge support from the Novo Nordisk Foundation Data Science Research Infrastructure 2022 Grant: A high-performance computing infrastructure for data-driven research on sustainable energy materials, Grant no. NNF22OC0078009.
\end{acknowledgments}

\def\bibsection{\section*{\refname}} 

\bibliography{references}

\pagebreak

\section{Appendices}
\subsection{Minimization of the cost function}
The cost function, Eq.~(\ref{eq:loglike}) together with the entropy term Eq.~(\ref{eq:entropy}) gives the regularized cost function, which is conveniently written (up to an additive constant)
\begin{align}
     \cost^\text{reg}(\bmw_0,\bmK) = &\frac{1}{2} \text{Tr} (
     \bmY \bmS^{-1} \bmY) + \frac{1}{2} \log(\det(\bmS)) \nonumber\\
     &- \frac{1}{2}\lambda_S \log(\det(\bmK+\lambda_K)) \label{eq:reg_cost}
\end{align}
where we have defined the matrices $\bmY_{nm} = (\bar{y}_n - t_n)\delta_{nm}$ and $\bmS_{nm} = \sigma_n^2 \delta_{nm}$ with $\bar{y}_n = (\dm\bmw_0)_n$ and  $\sigma_n^2 = (\dm(\bmK+\lambda_k)\dm^T)_{nn}$.

The regularized cost function is quadratic in $\bmw_0$, and it is therefore straightforward to find that at the minimum point of the cost, we have
\begin{equation}
      \bm{w}_0 = \left( \bm{\Phi}^T\bm{\Sigma}^{-1}\bm{\Phi}\right)^{-1}
  \bm{\Phi}^T\bm{\Sigma}^{-1} \bm{t}. \label{eq:w0}
\end{equation}

In order to efficiently minimize the cost function numerically, we need the derivative with respect to $\bmK$. We find this by using two formulas for the derivatives of an invertible matrix $\bmA$ with respect to a parameter $\theta$:
\begin{gather}
    \frac{\partial}{\partial\theta} \bmA^{-1} = 
    -\bmA^{-1} \frac{\partial\bmA}{\partial\theta} \bmA^{-1}\\
    \frac{\partial}{\partial\theta} \log(\det(\bmA)) = 
    \text{Tr}\left (\bmA^{-1} \frac{\partial \bmA}{\partial\theta}\right )
\end{gather}

The result is
\begin{align}
    \frac{\partial \cost^\text{reg}}{\partial\bmK} 
    = &\frac{1}{2}
    \dm^T(\bmS^{-1} -\bmY \bmS^{-2}\bmY) \dm \nonumber \\
    &- \frac{1}{2} \lambda_S (\bmK+\lambda_K)^{-1}
\end{align}

The covariance matrix has to be positive (semi-) definite, and we enforce this through Cholesky factorization
\begin{equation}
    \bm{K} = \bm{C}\bm{C}^T,
\end{equation}
where $C$ is a lower triangular matrix to be determined by the minimization. We therefore need the derivative of the cost with respect to $\bmC$, which becomes
\begin{equation}
    \frac{\partial \cost^\text{reg}}{\partial\bmC} 
    = 2 \frac{\partial \cost^\text{reg}}{\partial\bmK} \bmC.
\end{equation}

In practice, we seek solutions where the diagonal of $\bmC$ is positive, and we do this by writing the diagonal elements as squares of new variables $C_{ii} = c_i^2$ and use the chain rule $\frac{\partial \cost^\text{reg}}{\partial c_i}=2\frac{\partial \cost^\text{reg}}{\partial C_{ii}}c_i$.

The optimization is performed numerically with 100 different starting values for $\bm{C}$ to ensure that we reach a proper minimum.

\subsection{Calculated values for optimal parameters and covariance matrices}
We are considering two different sets of data. In the first situation, we only include atomization energies (Figs.~\ref{fig:atomization_PBE} and \ref{fig:atomization}). Because of the five-fold cross-validation, we are actually considering five different UAFDs. However, if we include all data points, the following parameters are obtained by minimizing the cost function:  $(w_1, w_2, w_3, w_4) = (4.69,-1.45,-2.22,1.71)$, and
\begin{equation}
\tilde{\bmK} =
\begin{pmatrix}
    11.05&-7.43&-17.14&7.83\\
    -7.43&5.47&14.24&-6.13\\
    -17.14&14.24&43.67&-17.65\\
    7.83&-6.13&-17.65&7.33
\end{pmatrix}
\end{equation}
In the case with simultaneous optimization of four different properties (Fig.~\ref{fig:Uncertainty_estimate_4db}) the optimal values, if all the data points are used, are $(w_1, w_2, w_3, w_4) = (-1.73,-0.11,1.64,-1.27)$, and
\begin{equation}
\tilde{\bmK} =
\begin{pmatrix}
    2.91&-0.25&-1.73&1.39\\
    -0.25&0.04&0.05&-0.04\\
    -1.73&0.05&2.98&-1.83\\
    1.39&-0.04&-1.83&1.25
\end{pmatrix}
\end{equation}

\end{document}